# Can scaling analysis be used to interpret the anti-parity-time symmetry in heat transfer?

Ying Li[1,2], Wei Li[3], and Cheng-Wei Qiu[1*]

[1] *Department of Electrical and Computer Engineering, National University of Singapore, Singapore 117583, Singapore*

[2] *Interdisciplinary Center for Quantum Information, State Key Laboratory of Modern Optical Instrumentation, ZJU-Hangzhou Global Science and Technology Innovation Center, College of Information Science and Electronic Engineering, Zhejiang University, Hangzhou 310027, China*

[3] *Department of Electrical Engineering, Ginzton Laboratory, Stanford University, Stanford, California 94305, USA*

**Abstract**

In a previous work (Li et al. Science 364, 170) [1], we proposed a heat transfer system that preserves the anti-parity-time (APT) symmetry, and observe the rest-to-motion phase transition during the symmetry breaking. Recently, it was suggested (Zhao et al. arXiv:1906.08431) [2] that the behaviours of the system can be understood using scaling analysis based on the Péclet and Nusselt numbers (Pe and Nu). It was further proposed that there exists a third regime in the phase diagram in addition to the symmetric and symmetry broken phases. Although we appreciate the proposal to characterize the contributions of coupling, diffusion, and advection with dimensionless numbers, here we show that they do not help to predict or interpret the behaviours of the APT system. The dimensionless numbers do not provide enough details about the system to conclude that there is a motionless phase, a phase transition, to find the critical point, or to give the correct phase diagram with only two regimes.

---

[*] e-mail: chengwei.qiu@nus.edu.sg

**1. The governing equation**

The system under consideration is made of two identical solid rings as in Fig. 1a. The inner and outer radius of each ring are $R$ and $R + \delta R$. The thickness is $b$. The thermal conductivity, density, and heat capacity of each ring are $\kappa$, $\rho$, and $c$. The upper ring is rotating as angular speed $\Omega$, while the lower ring is rotating at $-\Omega$. There is an interface of thickness $d$ and thermal conductivity $\kappa_I$ between the two rings. We are interested in the temperature distributions $T_1(x,t)$ and $T_2(x,t)$ along the inner edges of the upper and lower rings, where $x$ is the coordinate along each edge and $t$ is time. Under certain assumptions (Supplementary Material in [1]), the governing equations for $T_1(x,t)$ and $T_2(x,t)$ can be written as

$$\frac{\partial T_1}{\partial t} = D\frac{\partial^2 T_1}{\partial x^2} - \Omega R\frac{\partial T_1}{\partial x} + \frac{h}{\rho c b}(T_2 - T_1)$$
$$\frac{\partial T_2}{\partial t} = D\frac{\partial^2 T_2}{\partial x^2} + \Omega R\frac{\partial T_2}{\partial x} + \frac{h}{\rho c b}(T_1 - T_2)$$

(1)

where $D = \kappa/\rho c$ is the diffusivity of each ring, and $h$ is a coefficient used to characterize the heat exchange between the two rings. For the convenience of discussion, we use the same definition of $h$ as in [2]. According to the analysis in the Supplementary Material of [1], $h = \kappa_I/d$. Note that in [2] only the governing equation for $T_1$ is explicitly written, where $T_2$ is replaced with $T_\infty$. We understand that $T_\infty$ is used to justify $h$ as the heat transfer coefficient. However, the case here is different that $h$ is the heat exchange between two components of the system, and $T_\infty$ is an unknown variable instead of a constant as in the standard definition. Here we use $T_2$ and explicitly write both governing equations for clarity.

Eq. (1) can be transformed to an eigenvalue problem if we consider the time harmonic evolution

$$T_1(x,t) = T_0 + A_1 e^{ikx-i\omega t}$$
$$T_2(x,t) = T_0 + A_2 e^{ikx-i\omega t} \tag{2}$$

where $T_0$, $A_1$, and $A_2$ are constants, $k = \pm n/R$ ($n = 1, 2, \ldots$) considering the periodicity of $x$, and real parts should be taken for physical temperature fields. The eigenvalue of $\omega$ indicates that there is a phase transition when $\Omega$ is increased passing the critical value $\Omega_{EP} = h/\rho cb$, where $\omega$ turns from a purely imaginary number to a complex number with a finite real part. This phase transition is induced by a spontaneous anti-parity-time (APT) symmetry breaking. The evolution of $T_1(x,t)$ and $T_2(x,t)$ show a clear rest-to-motion transition correspondingly.

We impose an initial condition

$$T_1(x,0) = T_2(x,0) = T_0 + A\cos(x/R) \tag{3}$$

and perform numerical simulations for the system in [1]. As show in Fig. 2, the temperature profiles will not move but only decay with time when $\Omega < \Omega_{EP}$ (Fig. 1b), while they both move and decay with time when $\Omega > \Omega_{EP}$ (Fig. 1c). It meets the experimental results in Fig. 4 of [1]. We thank the authors for pointing out a typo in Fig.4G-I of [1]. In the figure insets, all $\Omega t$ should be $\Omega t/2$. We purposely did that to show the detailed variations of the measured $T_1$ fields (red lines). Otherwise, as in Fig. 2D-F of [2], the measured results are not clearly shown. We feel sorry for the confusion caused because of the fact that we did not mark this clearly in the insets.

**2. The scaling analysis**

In [2] it is proposed to defined the Péclet and Nusselt numbers for the system as: Pe = $\Omega R^2/D$, Nu = $hR/\kappa$. Before we proceed, we would like to comment that the use of Nu may be inappropriate. Recall that the Nusselt number is the ratio of convective heat over conductive heat in fluid. Here, although the rings are moving, they are solids. The coefficient $h = \kappa_i/d$ is induced by the thermal contact between the rings and the interface, it is independent of the motion of the rings. Therefore, $h$ cannot be regarded as the convective heat transfer coefficient that is used to define Nu. We propose that the relevant dimensionless number is the Biot number Bi = $hb/\kappa$, which characterizes the ratio of thermal conduction at the interface and inside the ring [3]. Note that the characteristic length is $b$ instead of $R$.

In [2] a variable change is made to nondimensionalize the governing equation. We follow a similar procedure to rewrite Eq. (1) as

$$\frac{\partial T_1^*}{\partial t^*} = \frac{\partial^2 T_1^*}{\partial x^{*2}} - \text{Pe}\frac{\partial T_1^*}{\partial x^*} + \eta \text{Bi}(T_2^* - T_1^*)$$
$$\frac{\partial T_2^*}{\partial t^*} = \frac{\partial^2 T_2^*}{\partial x^{*2}} + \text{Pe}\frac{\partial T_2^*}{\partial x^*} + \eta \text{Bi}(T_1^* - T_2^*) \quad (4)$$

where

$$T_1^* = (T_1 - T_0)/A, T_2^* = (T_2 - T_0)/A,$$
$$x^* = x/R, t^* = tD/R^2, \eta = R^2/b^2 \quad (5)$$

Unlike in [2], we use the constant term $T_0$ and the amplitude $A$ in the initial condition Eq. (3), instead of the reference temperature $T_\infty$ (which is just $T_2$ for the first equation and $T_1$ in the second) and the maximum temperature $T_{max}$, since both $T_\infty$ and $T_{max}$ vary with time. Note that Eq. (3) of [2] cannot be derived from the variable change. The correct version should have two equations as in Eq. (4), since the fields are coupled and must be solved together. Since we are using the Biot number, the definition of $\eta$ has also been changed.

A central claim of [2] is that by simply comparing Pe and $\eta$Bi ($\eta$Nu therein), one can understand the phase transition and obtain a phase diagram for the system. The authors claim that this is a conventional heat transfer framework. However, to the best of our knowledge, no theoretical frameworks have such predictions. First of all, there is no reason to use $\eta$Bi instead of Bi, which has concrete physical meaning. It seems that $\eta$Bi is proposed because it is the coefficient of the heat exchange term in Eq. (4), and because Pe = $\eta$Bi gives the critical value $\Omega_{EP} = h/\rho cb$.

More importantly, the dimensionless numbers only characterize the scales of certain physical quantities. For example, when Pe is much smaller than Bi, it is expected that the heat exchange between rings is dominant over the advection effect. However, it does not indicate that there should be a motionless phase or a phase transition with discontinuity at Pe = Bi (or Pe = $\eta$Bi). The proposed critical point at Pe = $\eta$Bi lacks physical insight and is based on the solutions given in [1]. It could lead to erroneous predictions when details of the system are different from that in [1].

For example, consider a slightly different system following

$$\begin{aligned}
\frac{\partial T_1}{\partial t} &= D\frac{\partial^2 T_1}{\partial x^2} - \Omega R\frac{\partial T_1}{\partial x} + \frac{h}{\rho cb}(T_\infty - T_1) \\
\frac{\partial T_2}{\partial t} &= D\frac{\partial^2 T_2}{\partial x^2} + \Omega R\frac{\partial T_2}{\partial x} + \frac{h}{\rho cb}(T_\infty - T_2)
\end{aligned} \quad (6)$$

where $T_\infty$ is a constant. Using a variable change $T_1^* = (T_1 - T_\infty)/A$ and $T_2^* = (T_2 - T_\infty)/A$, we have

$$\frac{\partial T_1^*}{\partial t^*} = \frac{\partial^2 T_1^*}{\partial x^{*2}} - \text{Pe}\frac{\partial T_1^*}{\partial x^*} - \eta \text{Bi} T_1^*$$
$$\frac{\partial T_2^*}{\partial t^*} = \frac{\partial^2 T_2^*}{\partial x^{*2}} + \text{Pe}\frac{\partial T_2^*}{\partial x^*} - \eta \text{Bi} T_2^* \quad (7)$$

Now the governing equation for $T_1$ looks exactly the same as the Eq. (3) in [2] (Eq. (3) in [2] has a positive sign in front of the third term which must be a typo, because it is unphysical that the temperatures will deviate from $T_\infty$). The system is actually different from the APT system in Fig. 1a. The two rings are in contact with an interface at constant temperature $T_\infty$ (Fig. 2a), so they are basically decoupled. It is easy to see that there is only one phase in such a system. The dependence of its behavior on Pe or Bi is smooth, and even for small Pe the profiles are still moving, just at lower speed (Fig. 2b and c). If the method used in [2] was indeed valid, one should expect a motionless phase when Pe < $\eta$Bi based on a governing equation of the same form.

In fact, when the two numbers are of the same order, one cannot make any prediction simply by comparing them without considering the solutions. As another example, we consider an initial condition different from Eq. (3)

$$T_1(x,0) = T_2(x,0) = T_0 + A\cos(2x/R) \quad (8)$$

which correspond to the solutions with $k = \pm 2/R$. In this case, the critical value becomes $\Omega_{EP} = h/2\rho cb$, or Pe = $\eta$Bi/2 according to the eigenvalue. However, the governing equation remains the same, so according to the method in [2], the critical point should still be Pe = $\eta$Bi. On the contrary, the numerical results in Fig. 3a and b show distinct phases for Pe < $\eta$Bi/2 and Pe > $\eta$Bi/2 (but still Pe < $\eta$Bi).

## 3. The diffusion regime in [2]

Finally, it was proposed in [2] that there are three regimes in the phase diagram of the system: coupling dominant (Pe < $\eta$Bi, $\eta$Bi > 1), advection dominant (Pe > $\eta$Bi, Pe > 1), and diffusion dominant (Pe < 1, $\eta$Bi < 1) regimes. This division is questionable, because there must be discontinuity at the interfaces between distinct phase regimes. However, the system behaves perfectly smooth when Pe or $\eta$Bi passes 1, nothing special happens at Pe = 1 or $\eta$Bi = 1 according to the solutions. Also, the coupling and advection regimes are still distinguishable in the diffusion regime.

Indeed, when Pe and $\eta$Bi are both much smaller than 1, the temperature fields will decay quickly. However, in [1] we are considering the transient states, not the steady states, so no matter how fast the fields decay, one can always measure the evolution of them and determine whether they are moving or not. In the long run, the system will become isothermal irrespective of its parameters. It is inappropriate for [2] to claim isothermal field as the characteristic of the diffusion regime. The numerical results in Fig. 4 confirm that there is no fundamental difference between Pe < $\eta$Bi < 1 (Fig. 4a) and Pe < 1 < $\eta$Bi (Fig. 4b), or $\eta$Bi < Pe < 1 (Fig. 4c) and $\eta$Bi < 1 < Pe (Fig. 4d). On the other hand, there are still two distinct phases in Pe < 1, $\eta$Bi < 1 as shown in Fig. 4a and Fig. 4c. Thus, the diffusion regime defined in [2] is invalid.

As a conclusion, we appreciate the proposal to nondimensionalize the heat transfer equations for the APT system in [2]. However, the dimensionless numbers can give very limited predictions about the behaviors of the system. The symmetry protected motionless phase, the phase transition, and the critical point cannot be inferred by simply comparing them. Forcefully using them may even lead to inaccurate predictions about the critical point and the phase diagram. Nevertheless, for more complex systems, scaling analysis could be helpful for qualitative comparisons. We encourage more in-depth derivations to clarify the roles of the dimensionless numbers in the full solutions.

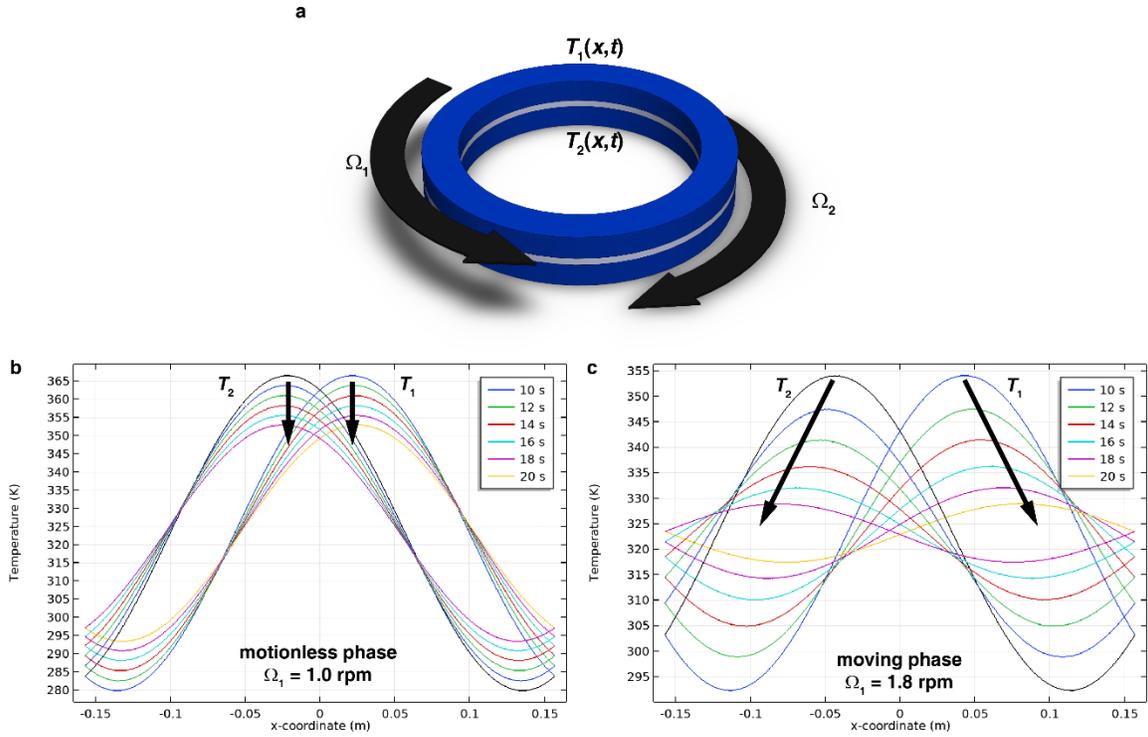

Fig. 1. **a**, Schematic for the APT system. **b**, For $\Omega < \Omega_{EP}$. The temperature profiles remain motionless. **c**, For $\Omega > \Omega_{EP}$. The temperature profiles are moving.

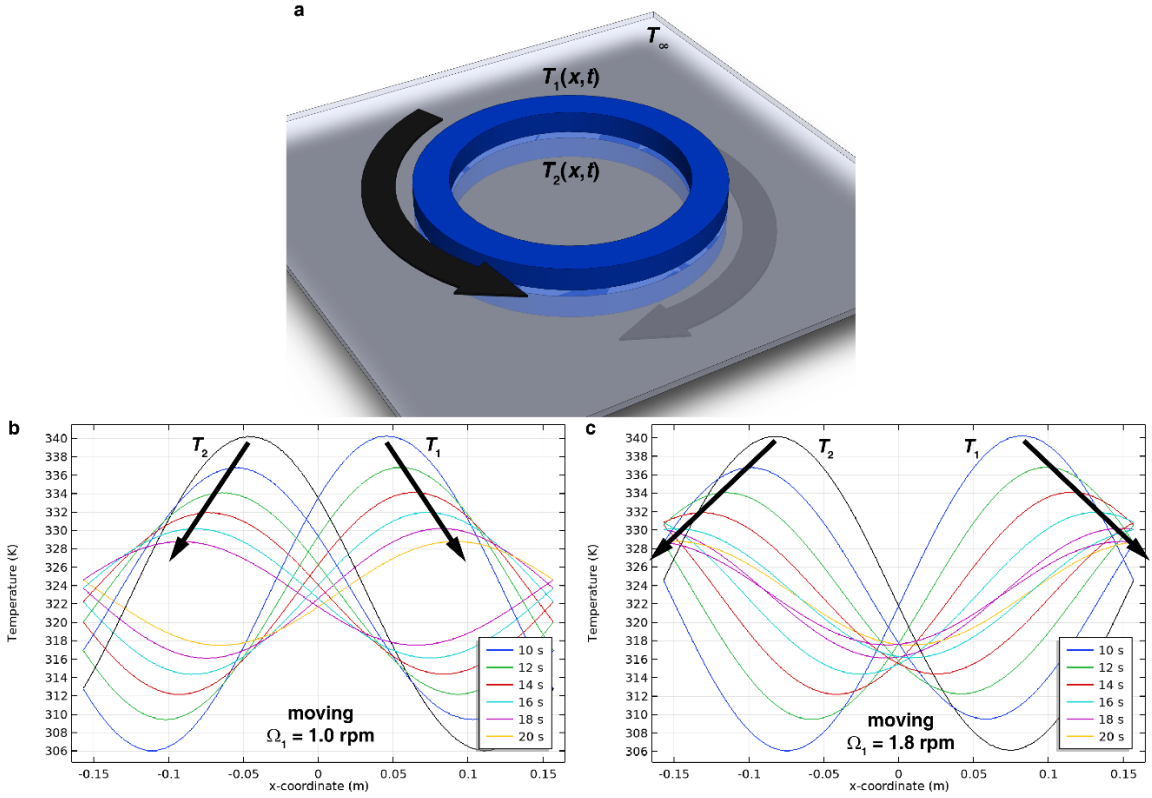

Fig. 2. **a**, Schematic for the system following Eq. (6). **b**, For $\Omega < \Omega_{EP}$ (Pe < $\eta$Bi), The temperature profiles are moving. **c**, For $\Omega > \Omega_{EP}$ (Pe > $\eta$Bi), The temperature profiles are moving.

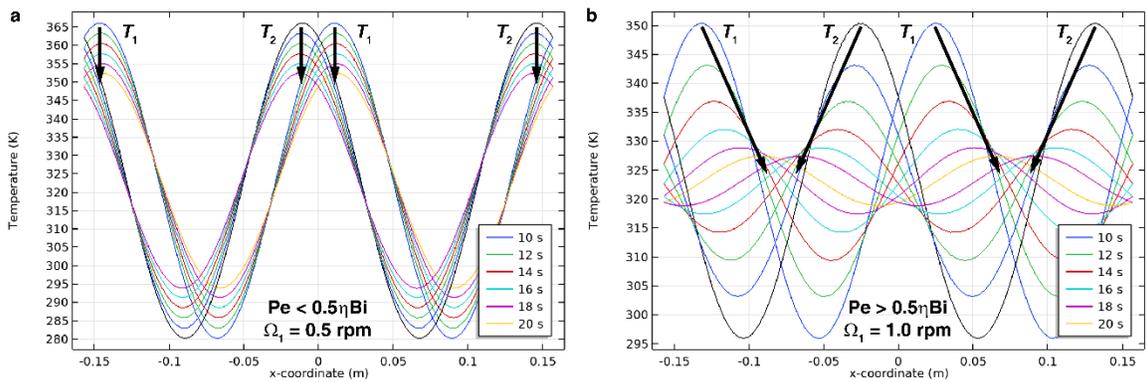

Fig. 3. A different initial condition. **a**, Temperature profiles remain motionless for Pe < 0.5$\eta$Bi. **b**, Temperature profiles are moving for Pe > 0.5$\eta$Bi (but Pe < $\eta$Bi).

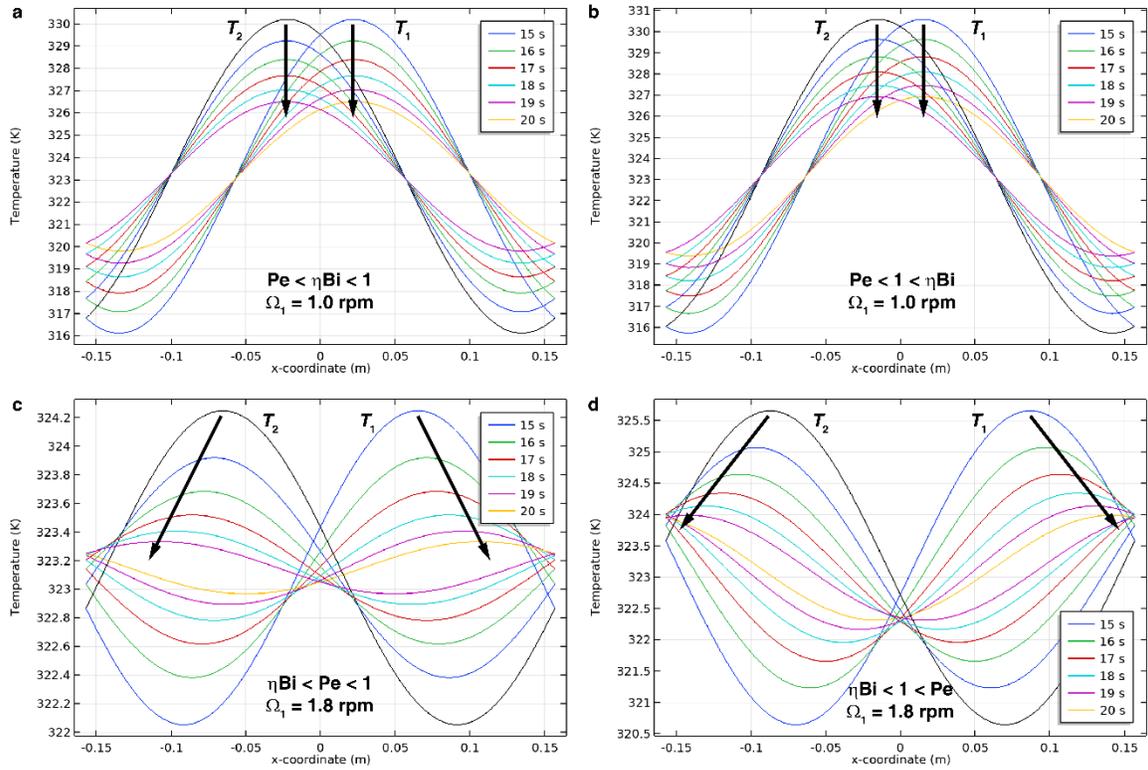

Fig. 4. **a**, Temperature profiles remain motionless for Pe < $\eta$Bi < 1. **b**, Temperature profiles remain motionless for Pe < 1 < $\eta$Bi. **c**, Temperature profiles are moving for $\eta$Bi < Pe < 1. **d**, Temperature profiles are moving for $\eta$Bi < 1 < Pe.